\documentclass[twocolumn,twoside,slac_two]{revtex4}
\usepackage{graphicx}
\usepackage{fancyhdr,natbib}
\begin{document}

\title{2FHL: The second Catalog of hard {\it Fermi}-LAT sources}

%

\author{M. Ajello, A. Dom\'{i}nguez}
\affiliation{Department of Physics and Astronomy, Clemson University, Kinard Lab of Physics, Clemson, SC 29634-0978, USA}
\author{S. Cutini, D. Gasparrini}
\affiliation{Agenzia Spaziale Italiana (ASI) Science Data Center, I-00133 Roma, Italy and
INFN, Sezione di Perugia,  I-06123 Perugia, Italy}
\affiliation{on behalf of the {\it Fermi}-LAT Collaboration}




\begin{abstract}
The {\it Fermi} Large Area Telescope (LAT) has been routinely gathering science data since August 2008, surveying the full sky every three hours. The first Fermi-LAT catalog of sources detected above 10 GeV (1FHL) relied on three years of data to characterize the $>$10 GeV sky. The improved acceptance and point-spread function of the new Pass 8 event reconstruction and classification together with six years of observations now available allow the detection and characterization of sources directly above 50 GeV. This closes the gap between ground-based Cherenkov telescopes, which have excellent sensitivity but small fields of view and short duty cycles, and all-sky observations at GeV energies from orbit. In this contribution we  present the second catalog
of hard Fermi-LAT sources detected at $>$50\,GeV.

\end{abstract}

\maketitle

\thispagestyle{fancy}


\section{Introduction}
\label{sec:intro}

The Large Area Telescope (LAT) on board the \emph{Fermi} satellite has been efficiently surveying
 the GeV sky detecting over 3000 sources in just
four years of exposure (see the 3FGL catalog, \cite{3FGL}).
These sources are detected in the 0.1--300\,GeV band and given {\it Fermi}'s
peak sensitivity  at $\sim$1\,GeV are representative of the GeV sky.
On the other hand, 
Cherenkov telescopes, with their good angular resolution and excellent
point-source sensitivity have been exploring, due to their limited
field of views, small patches of the $>50$\,GeV sky\footnote{VERITAS, H.E.S.S and MAGIC
have successfully lowered, in recent years, their low energy threshold 
and have started exploring the sub-100\,GeV band.}.
In the effort to fill the gap, the LAT-collaboration released a catalog
of sources detected, in 3\,years, at $>10$\,GeV (so called 1FHL  catalog, \cite{1FHL}).

Recently a new event reconstruction and characterization analysis (known as Pass 8, \cite{atwood13}) has been developed by the \emph{Fermi}-LAT collaboration. 
Pass~8 significantly improves the background rejection, point-spread function (PSF), effective area of the LAT and helps  understanding its systematic uncertainties. All these impressive improvements lead to a significant increase of the LAT sensitivity (Atwood et al., 2013a,b). Furthermore, these improvements are specially significant at $E>50$~GeV with an increase in the acceptance of $\gtrsim25$\,\% 
and an improvement in the PSF by a factor  between 
20\,\% at 50\,GeV and 50\,\% at 500\,GeV.
At these high energies, because of the almost lack of background, the sensitivity of 
{\it Fermi}-LAT  improves almost linearly with time as it should in a 
photon-limited regime (as opposed to a background-limited regime where the
sensitivity improves with the square root of exposure time).

Taking advantage of the improvements delivered by Pass~8, we are preparing
 an all-sky catalog of sources detected at $E>50$~GeV in $\sim$6\,years of data.
These sources will constitute the second catalog of hard {\it Fermi}-LAT sources (2FHL).
This proceeding shows that the 2FHL catalog provides a view of the high-energy
sky that is complementary to that of the 3FGL catalog and has the potential
to allow for unprecedented broad band studies of the SED of old and newly discovered sources
and to  increase the efficiency of the searches
of current Cherenkov telescopes.

%
%

\section{The 2FHL Catalog}
\label{sec:cat}
\begin{figure*}[ht!]
  \begin{center}
  \begin{tabular}{c}
  	 \includegraphics[scale=0.7,clip=True,trim=0 50 0 0]{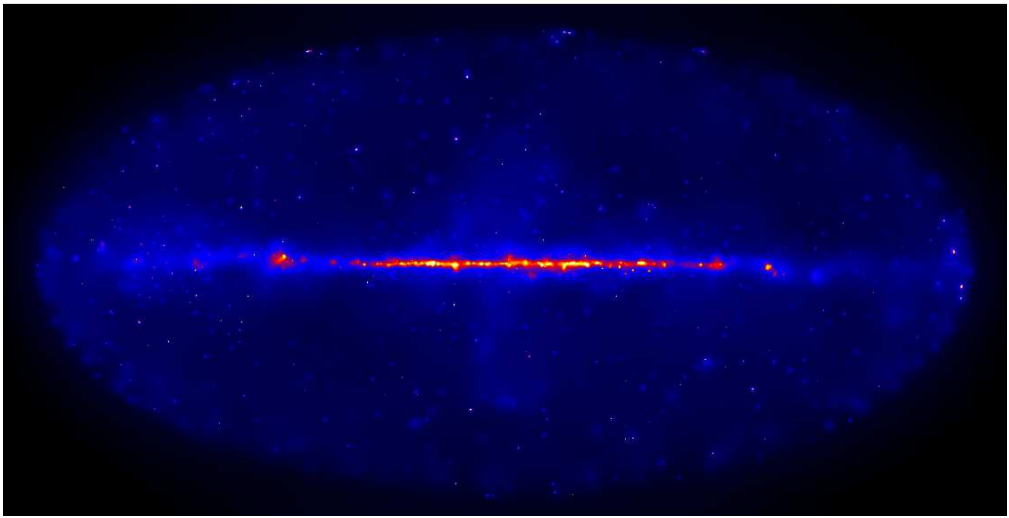} 
\end{tabular}
  \end{center}
\caption{Adaptively smoothed count map, in Galactic coordinates, at $>$50\,GeV. 
\label{fig:map}}
\end{figure*}

In about 6\,years of exposure,
{\it Fermi}-LAT has detected approximately 55000 photons (belonging to the P8 source
class) all-sky at $>$50\,GeV. 
The preliminary all-sky map in Fig.~\ref{fig:map} shows that {\it Fermi}-LAT
observes large scale diffuse emission in the direction of our Galaxy
and coincident with the so-called {\it Fermi} bubbles \citep{su2010,lat_bubbles} 
as well as many point-like sources.

The analysis to detect sources is performed similarly to the other {\it Fermi}-LAT
catalogs. The first step comprises the detection of source candidates (so called
seeds) as fluctuations above the background. The sky is then divided into
region of interests (ROIs), for which a sky model is built including all point sources
in the ROI and also the Galactic and isotropic diffuse models \cite{casandjan15}.
This model is fitted to the data via a standard maximum-likelihood unbinned algorithm.
The fit is typically repeated twice and in between the two fits the source position
is optimized using standard {\it Fermi} tools\footnote{In this case
gtfindsrc was used, see http://fermi.gsfc.nasa.gov/ssc/data/analysis/software.}.

Once a best fit has been found for a given ROI, the spectra of all sources
are generated in three logarithmic energy bins from 50\,GeV to 2\,TeV.

The 2FHL catalog comprises (preliminarily) $\sim$350 sources detected and characterized
exclusively at $>$50\,GeV.
For comparison, $\sim$145 are the known Very 
High-energy (VHE) sources reported in
the TeVCat\footnote{http://tevcat.uchicago.edu}. 
The 2FHL thus represents a leap forward for the study and characterization of
the VHE sky.
It is interesting to note that 2FHL sources are selected 
on the basis of their average flux and thus the 2FHL catalog may be considered an unbiased census of the
VHE sky. A preliminary association shows that $\sim$70 2FHL sources  are detected
in TeVCat as well and that the 2FHL comprises $\sim$100 sources that were not
detected in either the 1FHL or TeVCat.

Of all sources detected in the 2FHL, blazars (or blazar-like
objects) represent $\sim$75\,\%,
while unassociated sources and Galactic sources make up the rest.

\subsection{Angular and Position Resolution}
\begin{figure}[t]
  	 \includegraphics[scale=0.4,clip=True,trim=0 50 0 0]{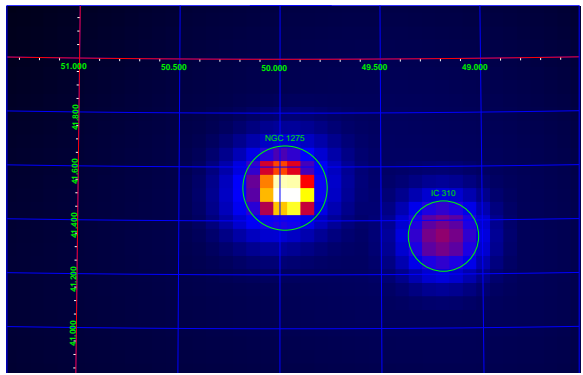} 
\caption{Adaptively smoothed count map of the region around NGC 1275 and IC 310
separated by roughly 0.6\,deg.
\label{fig:ngc1275}}
\end{figure}

Pass 8 improves the PSF of the LAT at all energies.
Above 50\,GeV the PSF has  a 68\,\% containment radius of $\sim$0.1$^{\circ}$
and remains constant with energy.
Such PSF, not dissimilar from the one of Cherenkov telescopes, allows
{\it Fermi}-LAT to localize sources with an average precision of 4$'$ at 95\,\% 
confidence. Fig.~\ref{fig:ngc1275} shows that {\it Fermi}-LAT can easily separate nearby
sources like it is the case for NGC 1275 and IC 310. 
However, such resolution is most useful in the plane of the Galaxy, where it
helps to solve crowded regions and resolve extended  sources.

%
%

\subsection{Spectra}

\begin{figure}[t]
\hspace{-1cm}
\includegraphics[scale=0.28,clip=True,trim=0 0 0 0]{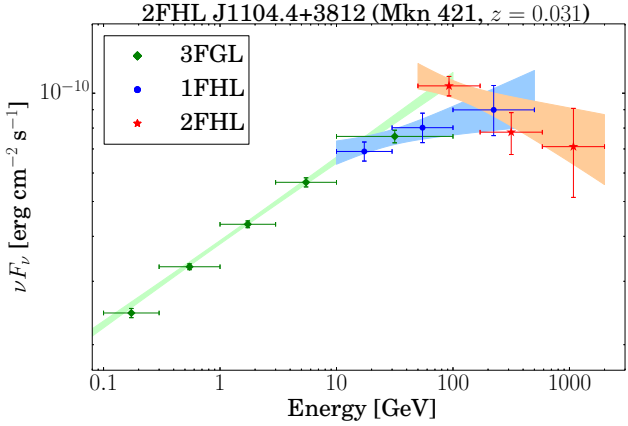} 
\caption{Preliminary spectrum of Mkn 421 in 2FHL together with data
from 1FHL and 3FGL. The three catalogs rely on different exposure times.
\label{fig:mrk421}}
\end{figure}

\begin{figure}[t]
  	 \includegraphics[scale=0.45,clip=True,trim=0 0 20 0]{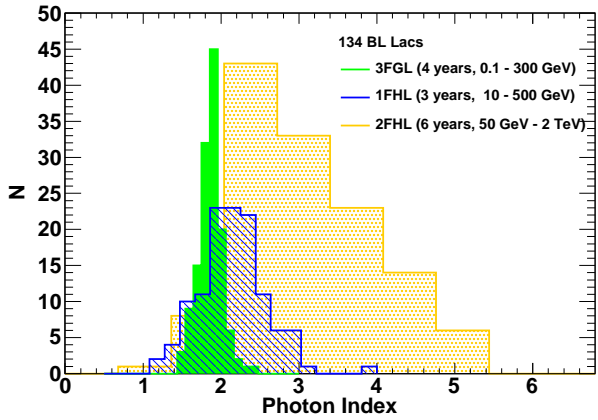} 
\caption{Distribution of the power-law photon indices for 134 BL Lacs
detected in the 3FGL, 1FHL and 2FHL catalogs. Note the softening
of the photon index when moving from lower energies (3FGL) to higher energies
(1FHL and 2FHL).
\label{fig:specdist}}
\end{figure}

The 2FHL catalog will report, for every source, 3 energy-bin spectra in the 
energy range 50\,GeV -- 2\,TeV. An example is reported, for Mrk 421, in
Fig.~\ref{fig:mrk421}. High synchrotron peaked (HSP) blazars, like Mkn 421,
are detected by {\it Fermi}-LAT, typically, as power-law sources
with a photon index of $\sim$1.8 (when integrated over the full
energy range as in the 3FGL). It is clear that above 50\,GeV 
(e.g. the 2FHL) {\it Fermi}-LAT samples
already the descending part of the high-energy peak of the spectral energy
distribution (SED) of such
sources and that the data from the 3FGL, 1FHL and 2FHL catalogs allow
us to characterize the emission at the peak of such sources rather well.
While Mkn 421 represents probably the best example, Fig.~\ref{fig:specdist}
shows that such conclusion holds, on statistical grounds, for most
BL Lacs detected by {\it Fermi}-LAT.

The 2FHL catalog comprises BL Lacs detected up to redshift $\sim$1.5.
The improved reconstruction and increased acceptance allow {\it Fermi}-LAT
to detect photons up to $\sim$2\,TeV (see e.g. Fig.~\ref{fig:mrk421}).
Both these aspects enable studies of the extragalactic background
light (EBL) which can absorb 
high-energy photons emitted from sources at cosmological distances
(EBL,\cite{gould66,fazio70,stecker92}). BL Lacs with substantial
high-energy emission at e.g. $>$100\,GeV are excellent probes of the 
EBL and have already been used with success to constrain the 
$\gamma$-ray opacity of the Universe \citep{ebl12,hess_ebl12,dominguez13a}.
We expect that the 2FHL, thanks to improved acceptance of
high-energy photons yielded by Pass 8, will enable accurate studies
of the EBL.

%
%

\section{Conclusions and Outlook}

The 2FHL catalog of {\it Fermi}-LAT sources detected at $>$50\,GeV
represents an unbiased census of the VHE sky. This work
 probes larger energies than any previous {\it Fermi}-LAT catalogs thanks to the
improved Pass 8 dataset. The view of the $\gamma$-ray sky
delivered by the 2FHL is complementary and different than
that of the (e.g.) 3FGL catalog. Indeed, we find that most extragalactic
sources are softer in the 2FHL than in the 3FGL, implying a peak
of their spectral energy distribution somewhere in the {\it Fermi} band.

The 2FHL catalog will comprise sources detected on the basis on their average flux.
Since 75\,\% of the detected sources
are blazars, the 2FHL will yield important information for the
generation of the high-energy part of the $\gamma$-ray background 
\citep{lat_egb2,ajello15}. It will also allow a first estimate
of the source count distribution of VHE sources acting
as a pathfinder for the surveys performed by the
upcoming Cherenkov Telescope Array \citep{dubus13}.

Finally, the good angular resolution achieved, thanks to Pass 8,
by {\it Fermi}-LAT at $>$50\,GeV will allow unprecedented
studies of the Galaxy allowing to resolve crowded regions
as well as new extended sources.  We envision that this aspect of
2FHL will act as a lower energy counterpart of the H.E.S.S.
Galactic plane survey \citep{carrigan2013} and the survey
carried out by HAWC \citep{westerhoff14}.

\bigskip 
\begin{acknowledgments}

The Fermi LAT Collaboration acknowledges generous ongoing support from a number of agencies and institutes that have supported both the development and the operation of the LAT as well as scientific data analysis. These include the National Aeronautics and Space Administration and the Department of Energy in the United States, the Commissariat à l'Energie Atomique and the Centre National de la Recherche Scientifique / Institut National de Physique Nucléaire et de Physique des Particules in France, the Agenzia Spaziale Italiana and the Istituto Nazionale di Fisica Nucleare in Italy, the Ministry of Education, Culture, Sports, Science and Technology (MEXT), High Energy Accelerator Research Organization (KEK) and Japan Aerospace Exploration Agency (JAXA) in Japan, and the K. A. Wallenberg Foundation, the Swedish Research Council and the Swedish National Space Board in Sweden.
Additional support for science analysis during the operations phase is gratefully acknowledged from the Istituto Nazionale di Astrofisica in Italy and the Centre National d'Etudes Spatiales in France.
\end{acknowledgments}

\bigskip 
\bibliography{/Users/majello/Work/Papers/BiblioLib/biblio.bib}

\end{document}